\documentclass[
prx,
a4paper,     
twocolumn,   
noshowpacs,  
showkeys,  
notitlepage, 
amsfonts,    
amsmath,     
amssymb      
]{revtex4-2}

\usepackage[dvipdfmx]{graphicx}
\usepackage{dcolumn}
\usepackage{bm}
\usepackage{times}
\usepackage[usenames]{color}
\usepackage[normalem,normalbf]{ulem}

\begin{document}

\title{Detecting partial synchrony in a complex oscillatory network using pseudo-vortices}

\author{Yasuhiro Yamada$^\dagger$}
\email{yshr.yamada@ntt.com}
\thanks{$^\dagger$ These authors contributed equally to this work.}
\author{Kensuke Inaba$^\dagger$}
\affiliation{NTT Basic Research Laboratories, NTT Corporation, 3-1 Morinosato Wakamiya, Atsugi, Kanagawa, 243-0198, Japan}

\date{\today}

\begin{abstract}
Partial synchronization is characteristic phase dynamics of coupled oscillators on various natural and artificial networks, which can remain undetected due to the complexity of the systems.
With an analogy between pairwise asynchrony of oscillators and topological defects, {\it i.e.,} vortices, in the two-dimensional XY spin model, we propose a robust and data-driven method to identify the partial synchronization on complex networks.
The proposed method is based on an integer matrix whose element is pseudo-vorticity that discretely quantifies asynchronous phase dynamics in every two oscillators, which results in graphical and entropic representations of partial synchrony.
As a first trial, we apply our method to 200 FitzHugh-Nagumo neurons on a complex small-world network.
Partially synchronized chimera states are revealed by discriminating synchronized states even with phase lags. Such phase lags also appear in partial synchronization in chimera states. Our topological, graphical, and entropic method is implemented solely with  measurable phase dynamics data, which will lead to a straightforward application to general oscillatory networks including neural networks in the brain.
\end{abstract}


\maketitle

\newpage

\section{Introduction}
Synchronization is a universal phenomenon in a system of interacting oscillators whose relative phases are fully locked in the whole system~\cite{pikovskij_synchronization_2007}.
In recent decades, partial synchronization of oscillators has also been recognized as ubiquitous and important in broad research fields through extensive chimera studies originating from seminal theoretical papers~\cite{kuramoto_coexistence_2002,abrams_chimera_2004}.
At the beginning, chimera states denoted self-sustained heterogeneous states composed of synchronous and asynchronous oscillators with a symmetry breaking in homogeneous oscillators, {\it e.g.} a ring of identical phase oscillators~\cite{kuramoto_coexistence_2002,abrams_chimera_2004}.
Chimera states are found in experiments on optical~\cite{hagerstrom_experimental_2012,larger_laser_2015,hart_experimental_2016,shena_turbulent_2017,brunner_spatio-temporal_2018,makinwa_experimental_2022}, chemical~\cite{tinsley_chimera_2012,nkomo_chimera_2013,schmidt_coexistence_2014,wickramasinghe_spatially_2013,totz_spiral_2018}, mechanical~\cite{martens_chimera_2013,kapitaniak_imperfect_2015}, and  electronic~\cite{larger_virtual_2013,rosin_transient_2014,gambuzza_experimental_2014,cao_cluster_2018} oscillators.
Recently, the term is generally used for partially synchronized states appearing in various fields on physics, chemistry, biology, engineering, and social networks, irrespective of the presence of the symmetry breaking~\cite{panaggio_chimera_2015,scholl_synchronization_2016,majhi_chimera_2019,wang_brief_2020,zakharova_chimera_2020,parastesh_chimeras_2021}.
In fact, partially synchronized chimera-like phenomena possibly have significant influence on the  functionalities and health of brain neurons~\cite{mukhametov_interhemispheric_1977,rothkegel_irregular_2014,mascetti_gian_gastone_unihemispheric_2016,rattenborg_evidence_2016,asher_connectivity_2021}, on heart rhythm~\cite{cherry_visualization_2008}, on power-grid networks~\cite{motter_spontaneous_2013}, and on cluster synchronization on irregular graphs ~\cite{kaneko_clustering_1990,bergner_remote_2012,nicosia_remote_2013,ashwin_weak_2015,sorrentino_complete_2016,schaub_graph_2016,bick_isotropy_2017,cao_cluster_2018,siddique_symmetry-_2018}, which  have inherent inhomogeneity on complex networks.

Our motivation is to quantify the expanding partial synchrony on complex networks in a robust and data-driven way.
In the original chimera studies on a regular ring, two kinds of synchrony based on frequency and phase were used to clarify the chimera states~\cite{kuramoto_coexistence_2002,abrams_chimera_2004}.
Frequency synchronization indicates that synchronized oscillators possess the same oscillation frequency as a result of entrainment.
A distribution of mean or instantaneous frequency is often used to detect the partial frequency synchrony,
where the coexistence of synchronized oscillators with the same frequency and desynchronized ones with distributed frequency indicates a chimera state~\cite{kuramoto_coexistence_2002,abrams_chimera_2004, nkomo_chimera_2013,martens_chimera_2013,wickramasinghe_spatially_2013,rosin_transient_2014,totz_spiral_2018,makinwa_experimental_2022}.
While the frequency synchrony qualitatively reveals the chimera states with a look at the distribution,  phase synchrony has been utilized to quantify chimeras sufficiently.
A region of oscillators is obviously synchronized with spatial coherence when the oscillators are almost in-phase at a time, which can be easily checked with the phase snapshot~\cite{kuramoto_coexistence_2002,abrams_chimera_2004,hagerstrom_experimental_2012,nkomo_chimera_2013,schmidt_coexistence_2014,martens_chimera_2013,larger_virtual_2013,kapitaniak_imperfect_2015,gambuzza_experimental_2014,rosin_transient_2014,larger_laser_2015,brunner_spatio-temporal_2018,shena_turbulent_2017,totz_spiral_2018,makinwa_experimental_2022}.
The spatial synchrony is quantified by the distribution of the local order parameter $r_x=|\sum_{x'} g(x,x')\exp(i\theta_{x'}))|$ where $g(x,x')$ determines the local vicinity of an oscillator at a position $x$~\cite{kuramoto_coexistence_2002,abrams_chimera_2004,tinsley_chimera_2012,nkomo_chimera_2013,wickramasinghe_spatially_2013,totz_spiral_2018}.
The order parameter is also utilized for analytical investigations of chimera states~\cite{kuramoto_coexistence_2002,abrams_chimera_2004,abrams_solvable_2008,allefeld_detecting_2007,bialonski_identifying_2006,kotwal_connecting_2017}.
To quantify the in-phase synchrony more clearly, Kemeth {\it et al.} introduced the local curvature or the pairwise Euclidean distance of oscillator phases, and proposed chimera measures with a threshold for the spatial coherence to classify the chimera states~\cite{kemeth_classification_2016}, which has been successively used to detect experimental chimeras ~\cite{shena_turbulent_2017,makinwa_experimental_2022}.
The frequency and in-phase synchronies are interchangeable in original chimera studies, where the partially frequency-synchronized oscillators are almost in-phase~\cite{kuramoto_coexistence_2002,abrams_chimera_2004}. 
For other chimera measures based on frequency and phase synchrony, see review~\cite{parastesh_chimeras_2021}.

However, we sometimes face difficulties in quantifying the partial synchrony apart from the original context.
For instance, quantitative indices of the spatial phase coherence can miss the synchrony that occurs remotely on an irregular graph~\cite{bergner_remote_2012} or with twisted/spiral waves ~\cite{tsimring_repulsive_2005,wiley_size_2006,cherry_visualization_2008,girnyk_multistability_2012,totz_spiral_2018}.
Although the remotely synchronized clusters can be predicted by local symmetries of the graph~\cite{nicosia_remote_2013}, we need to know the precise network structure and the exact equation of motion for oscillators, which are not necessarily accessible in actual experiments.
With the expanding interest in partial synchrony, we will obtain more data on partial synchrony in noisy experiments of oscillators on complex networks with unknown details. We will analyse them and try to utilize the functionality of partial synchrony for applications, like for optimization in an opto-electric experiment~\cite{inagaki_collective_2021}.
Developing a robust concept of synchrony against noises only based on measurable variables is, therefore, desirable for computer-friendly quantitative identification of the expanding partial synchrony in this data-driven era.

In this paper, we propose a discrete analogue of frequency synchrony with an analogy between partial asynchrony and topological defects, {\it vortices}, in the two-dimensional XY spin model~\cite{kosterlitz_ordering_1973} to quantify the expanding partial synchrony clearly and robustly: no vortex in a space-time (1+1 dimensional) plane means synchrony.
Using the analogy, we discretely quantify the accumulated phase mismatch between every two oscillators during a period, which leads to an integer matrix $I$.
We note that the matrix can be calculated solely from oscillator phase dynamics data without knowledge of the network structure or theoretical model that describes the oscillator dynamics.
The element $I_{ij}$ is {\it pseudo-vorticity} that counts pairwise asynchrony between $i$th and $j$th oscillators. 
To quantify the partial synchronization in a whole system, we next introduce a graph from the matrix, named a {\it synchronization graph}, where synchronous oscillators are connected with each other irrespective of the original network structure. From the synchronization graph, we identify the clique clusters in which all oscillators are mutually synchronized by solving a clique problem of the graph. 
The size distribution of the clusters gives an entropic measure that quantifies the disorder of the synchronization, which becomes a useful measure of chimera states. 
The partially synchronized states are further classified by {\it frequency divergence} obtained from the pseudo-vorticity matrix. 
Numerical simulations confirm that these measures identify different phases in the parameter space of FitzHugh-Nagumo (FHN) neurons~\cite{fitzhugh_mathematical_1955,nagumo_active_1962} on a small-world network~\cite{watts_collective_1998} with high clustering and a short path length. The FHN neuron is a brain-inspired neuronal oscillator, which has been used extensively in researches on chimera states on a one-dimensional ring~\cite{omelchenko_ringFHN_2013,gambuzza_experimental_2014, gambuzza_experimental_2014,semenova_brainFHN_2016}, empirical brain networks~ \cite{chouzouris_chimera_2018,ramlow_partial_2019,gerster_fitzhughnagumo_2020}, and networks with complex topology~\cite{omelchenko_ringFHN_2015,huo_brainFHN_2019,rontogiannis_chimera_2021}. Network small-worldness is one of the important features of actual neural networks ~\cite{watts_collective_1998}.
We emphasize that our method property identifies synchronization with a gradating phase lag and the related chimeras without additional human confirmation, which cannot be identified by previous chimera measures.

\section{Theoretical Framework}
A topological index can be useful for a robust classification of a system in general because it gives a measure of fundamental classification without information on the fine details of the system. A famous example is the topological defects in the two-dimensional $XY$ spin model, which characterize the low energy excitations in the system and account for the topological phase transition~\cite{kosterlitz_ordering_1973}. A topological defect is discretely quantified by the vorticity obtained from an integral of the spin phase along a closed path as shown in Fig. 1(a); $v_c=\frac{1}{2\pi}\oint_{C}\vec{\nabla} \theta(\vec{r})\cdot d\vec{r}=\frac{1}{2\pi}\oint_{C}d\theta$.

Similarly, we introduce an integer index that characterizes the partial synchronization. As a preparation for the following discussion, we consider a one-dimensional oscillator field $\theta(x,t)$ whose phase continuously changes along the continuous $x$-axis as well as along time. In this case we can define a topological quantity by an integral of the phase field as a function of time and the oscillator index on a closed path $C(x,t';t,t')$ as in Fig.1(b);
\begin{equation}
I(x,x';t,t')=\frac{1}{2\pi}\oint_{C(x,t;x,t')}d\theta(x,t),
\end{equation}
which represents an integer vortex strength generated by the phase field on the (1+1) dimension. The zero value $I(x,x';t,t')=0$ is necessary for the condition that the domain from $x$ to $x'$ is synchronized during the time evolution from $t$ to $t'$ with no accumulated vortex. Then, we can relate the partial synchronization with the integral.

\begin{figure}[tb]
\begin{center}
\includegraphics[width=8.8cm]{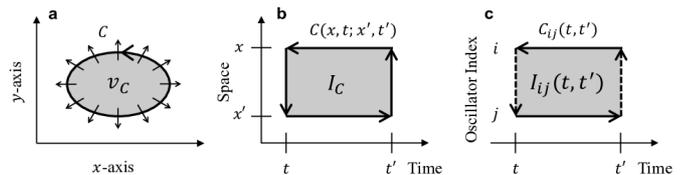}
\end{center}
\caption{Schematic illustrations of closed paths for calculating vorticity and pseudo-vorticity. (a) Vorticity $v_C$ bounded by a closed curve $C$ in two-dimensional phase field $\theta(x,y)$. (b) Vorticity $I_C$ bounded by a closed rectangular $C(t,t';x,x')$ in 1+1 dimensional phase field $\theta(t,x)$. (c) Pseudo-vorticity $I_{ij}(t,t')$ defined in between oscillator $i$ and $j$.}
\label{fig1}
\end{figure}

The integral of the phase field is, however, not well defined in the case of the discrete oscillators considered here because the phase may discontinuously change along the oscillator index with a phase ambiguity of the amount of $2\pi$.  
We here define an analogous index by dividing the closed path into the continuous changing parts along the time and discontinuous changing parts along the oscillator index as shown in Fig.~\ref{fig1}(c);
\begin{align}
I_{ij}(t,t')&\equiv\frac{1}{2\pi}\Big[\int_{\theta_j(t)}^{\theta_j(t')}d\theta_j(t)+\int_{\theta_i(t')}^{\theta_i(t)}d\theta_i(t)\notag\\
&\quad+\textrm{Arg}(e^{i(\theta_i(t')-\theta_j(t'))})+\textrm{Arg}(e^{i(\theta_j(t)-\theta_i(t))})\Big]
\end{align}
where the argument function $\textrm{Arg}(x)$ terms account for the phase difference between $i$th and $j$th oscillators at the initial and the final times $t$ and $t'$.
It is noteworthy that the index is an integer, $I_{ij}(t,t')=\lfloor\frac{1}{2}+\frac{\theta_{j}(t')-\theta_{i}(t')}{2\pi} \rfloor+\lfloor\frac{1}{2}+\frac{\theta_{i}(t)-\theta_{j}(t)}{2\pi} \rfloor$, that is associated with the pseudo-vortex strength inside the shaded box in Fig.~\ref{fig1}(c). The pseudo-vorticity is directly related to the phase difference between two oscillators accumulated during the time interval, $\varDelta\Phi_{ij}(t,t')\equiv\theta_j(t')-\theta_j(t)-\theta_i(t')+\theta_i(t)$, by the following inequality
\begin{equation}
\lfloor \frac{\varDelta\Phi_{ij}(t,t')}{2\pi}\rfloor\le I_{ij}(t,t')\le\lfloor \frac{\varDelta\Phi_{ij}(t,t')}{2\pi}\rfloor+1.
\label{eq:phase_relation}
\end{equation}
To see the physical meaning clearly, we consider a simple case where all oscillators have definite harmonic motions, $\theta_i(t)=2\pi f_i t +\phi_i$. In this case, the pseudo-vorticity holds the inequality  $\lfloor(f_j-f_i)(t'-t)\rfloor\le I_{ij}(t,t')\le \lfloor(f_j-f_i)(t'-t)\rfloor +1$, which clearly indicates that the integer element approximately gives the phase difference accumulated by the time-evolution, $I_{ij}(t,t')\sim(f_j-f_i)(t'-t)$.

For $n$ oscillators, we have an integer $n\times n$ matrix $I$ whose elements are given by the pseudo-vorticities. The {\it pseudo-vorticity matrix} satisfies a couple of symmetries. The matrix is almost antisymmetric, $I(t,t')=-I^{T}(t,t')$ in terms of the oscillator index, and is certainly antisymmetric in terms of the time variables, $I(t,t')=-I(t',t)$. It also satisfies the following triangle identity in time, $I(t_1,t_2)+I(t_2,t_3)+I(t_3,t_1)=O$ because of the continuous phase change along the time-evolution. Although a similar identity for the node indices does not hold in general, $I_{ij}(t,t')+I_{jk}(t,t')+I_{ki}(t,t')\neq 0$ due to the discontinuous phase change along the oscillator index.

It is noteworthy that the measurement time interval $\varDelta t:=|t'-t|$ is important to distinguish synchronization from desynchronization. Obviously, for $\varDelta t \to 0$, the pseudo-vorticity matrix trivially becomes the $n\times n$ null matrix, which seemingly indicates the synchronization for any case since we need a finite interval $\varDelta t$ to discriminate the desynchronization of two oscillators from the frequency synchronization. The minimum requirement of the time interval is approximately given by the minimum frequency difference $\varDelta f_{min}$ among all oscillators, $\varDelta t \approx |\varDelta f_{min}|^{-1}$. However, the minimum value is usually unclear before executing a measurement or a numerical simulation in actual cases, which means that a given measurement interval $\varDelta t$ determines a resolution of frequency synchrony. Then, we need to carefully check whether a given $\varDelta t$ is enough to distinguish synchronization from desynchronization or not by changing its value. Furthermore, $\varDelta t$ is also related to the fundamental trade-off between resolutions of timing (time) and frequency to detect a transient synchronization; a longer measurement interval $\varDelta t$ provides a finer resolution in frequency, while it also gives rise to a reduction in detectability to the time-dependence of synchronization.

The pseudo-vorticity matrix includes the information on pairwise synchrony of every two oscillators. To quantify the partial synchronization in a whole system, we introduce a graphical representation of the (partially) synchronized groups of oscillators with an adjacency matrix, $A$, whose elements are defined by
\begin{equation}
A_{ij}(t,t')=
\begin{cases}
1 & |I_{ij}(t,t')|\le c_{s}, i\neq j,\\
0 & \textrm{otherwise},
\end{cases}
\end{equation}
where $c_{s}$ is the synchrony criteria of pseudo-vorticity. 
From Eq.~\ref{eq:phase_relation}, $c_{s}=0$ and $c_{s}=1$ are sufficient and necessary conditions for $-2\pi\le\varDelta\Phi_{ij}(t,t')<2\pi$, respectively. Hereafter, we take $c_{s}=1$ to find the synchrony with the phase difference within $2\pi$. 
The adjacency matrix $A(t,t')$ is symmetric because the pseudo-vorticity matrix $I(t,t')$ is anti-symmetric.

Every two connected nodes in the {\it synchronization graph} with the adjacency matrix are pairwise synchronized. If there is a group of nodes connected completely, i.e., a clique, every two nodes in the group are synchronized, and the group or clique represents a partial synchronization in all the oscillators. Here we divide all the nodes into synchronized clique groups. At the first step, we find the largest clique in the synchronization graph, which is the maximum clique problem of the graph and a basic NP hard problem in computer science. Computing all cliques requires an exponential time with increasing graph size in general, but can be available for a relatively small size or sparse graph. In addition, we can obtain the approximated solution by solving the corresponding optimization problem. With a numerical algorithm, we (approximately) obtain the largest clique of the graph. Next, we find the next largest clique in the graph after removing the maximum clique. By repeating the procedure iteratively, we can identify all the cliques which need to cover the graph and their sizes, $s_1,\dots,s_m$, with $m$ being the number of cliques.
This procedure mathematically gives a partition of the oscillator set, which naturally induces an equivalence relation of partial synchrony on the oscillators.

From the clique sizes, we define the probability where a node belongs to the $j$th largest clique by
\begin{equation}
p_{j}=s_{j}/n\qquad\textrm{for}\quad j=1,\dots,m
\end{equation}
where $n$ is the total number of nodes. If a single clique can cover the graph with $p_1=1$, then the whole system is synchronized, where any node must belong to the unique maximum clique. On the other hand, if every node belongs to different cliques, the system is totally desynchronized with the uniform distribution, $p_1=p_2=...=p_n=1/n$. Therefore, the partiality of synchrony (the uniformness of the probability) is quantified by the entropy as
\begin{equation}
S_{sync}:=-\sum_{j=1}^{m}p_j\ln p_j
\label{eq:sync_entropy}
\end{equation}
At $S_{sync}=0$, the system seems to be synchronized with almost no vortex between every two nodes, while the system is completely desynchronized when the entropy has the maximum value, $S_{sync}=\log n$. In between them, the system comprises a mixture of several synchronized parts and desynchronized parts.
Then, we here refer the entropy to as {\it synchronization entropy}.

In addition to the entropic measures, we calculate a characteristic frequency by using the measurement time interval and the pseudo-vorticity matrix,
\begin{equation}
\varDelta f:=\frac{||I(t,t')||_{F}}{\sqrt{2}n\varDelta t},
\end{equation}
where $||A||_{F}$ denotes the Frobenius norm of a matrix $A$. We here refer to the frequency as {\it frequency divergence} because it physically means the typical frequency difference among oscillators. For instance, the frequency divergence is asymptotically equivalent to the standard deviation of the frequencies, $\varDelta f \approx \sqrt{\frac{1}{n}\sum_{i=1}^{n}(f_i-\bar{f})^2}$, for non-interacting harmonic oscillators and a long interval $\varDelta t$, where $f_i$ and $\bar{f}$ are the individual frequencies and the average frequency of oscillators. If the frequency divergence is less than the inverse of the measurement time interval, oscillators almost seem to have a single frequency in the measurement interval on average. Otherwise, the system includes oscillators with a variety of frequencies;
\begin{equation}
\varDelta f \varDelta t
\begin{cases}
\ll 1 & \textrm{single-frequency oscillators},\\
\gg 1 & \textrm{multi-frequency oscillators},
\end{cases}
\end{equation}
The frequency divergence does not identify partial synchronizations bbecome a useful tool and ecause a partially synchronized oscillatory system may, or may not have a difference in frequency on average. Nonetheless, it is helpful for classifying partial synchronizations further. 
For example, in Kuramoto and FHN oscillators on the ring structure, a desynchronized region in chimera states show frequency bands even though the original frequencies without interactions are unique~\cite{kuramoto_coexistence_2002,omelchenko_ringFHN_2013,omelchenko_ringFHN_2015}.
The frequency divergence can capture such a change in the frequencies caused by interactions.

For a completely synchronized system, the proposed entropic and frequency measures must be zero. Therefore, they can be regarded as "disorder parameters", which are the order parameters of the desynchronization quantified from different points of view. These disorder parameters are complementary to the ordinal order parameter of complete phase synchronization, $r=|\sum_j \exp(i\theta_j)|$. 

The proposed method is simply based on basic concepts of topology, graph theory, and information theory, which are well developed in their fields.
We can straightforwardly execute the method with a computational algorithm and raw phase dynamics data.
Note that Boltzmann-Gibbs-Shannon entropy used in Eq.~\eqref{eq:sync_entropy} is a choice to quantify the partiality of synchrony. Other possible choices are discussed in Appendix A.

\section{Application to FitzHugh-Nagumo neuronal oscillators on a small-world network}
As the first benchmark, we used the present synchronization measures to investigate the FHN model on a small-world network. The FHN model~\cite{fitzhugh_mathematical_1955,nagumo_active_1962} is a simplification of the Hodgkin–Huxley model~\cite{hodgkin_quantitative_1952} that describes neuron dynamics. It has been pointed out that the small-worldness appears in a variety of systems, such as actual neuronal networks~\cite{watts_collective_1998}. Thus, the following test case of the usage of our proposed method may extend to the empirical results obtained from actual brain experiments.

The dynamics of the FHN model is written as: 
\begin{equation}
\frac{1}{\tau}\frac{dv_i}{dt}=v_{i}-\frac{v_i^3}{3}-w_i+\cos\alpha\sum_{j}J_{ij}v_j+\sin\alpha\sum_j J_{ij}w_j
\end{equation}
\begin{equation}
\frac{dw_i}{dt}=a-bw_i+v_i+\cos\alpha\sum_{j}J_{ij}w_j-\sin\alpha\sum_j J_{ij}v_j,
\end{equation}
where $v$ and $w$ are membrane potential and recovery variables, respectively, and $\alpha$ is a correlation lag. A coupling matrix $J_{ij}$ is given by $J_{ij}=(K/N)a_{ij}$, where $K$ is coupling strength, $N$ is the number of sites, and $a_{ij}$ is the adjacency matrix of a small-world graph. Other parameters are set as $a=0.5$, $b=0$, and $\tau=0.05$. Small-world graph $a_{ij}$ is created by the rewiring scheme proposed in \cite{watts_collective_1998} by starting from $a_{ij}$ of a ring lattice of $N=200$ sites with $k=6$ edges. An obtained graph satisfies the characteristic of the small-world graph as follows. It has a clustering coefficient of 0.654, which is much greater than $k/n=0.05$, and characteristic path length of 6.06, which is on the same order as $\ln(N)/\ln(k)=2.30$, where $k/n$ and $\ln(N)/\ln(k)$ are the clustering coefficient and characteristic path length of the corresponding random graph. 

\begin{figure}[tb]
\begin{center}
\includegraphics[width=8.8cm]{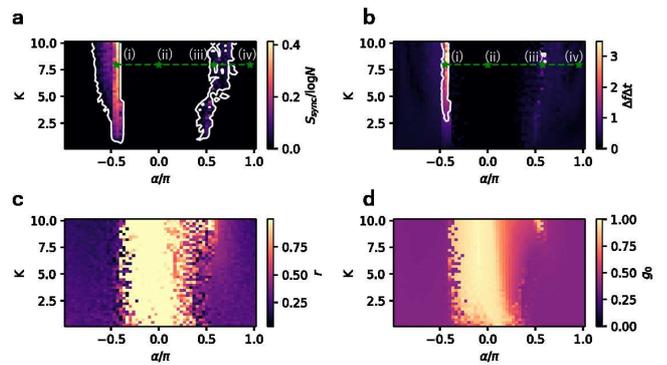}
\end{center}
\caption{False-color maps of synchronization measures of coupled FitzHugh-Nagumo neurons on a small-world network. (a), Synchronization entropy $S_{sync}$, (b), Frequency divergence $\varDelta f \varDelta t$, (c), conventional phase order parameter $r$, and (d), relative size of the coherent component $g_0$. The horizontal and vertical axes represent the correlation lag $\alpha$ and the coupling strength $K$, respectively.}
\label{fig2}
\end{figure}
Figure \ref{fig2}(a) and (b) show the synchronization entropy $S_{sync}$ and frequency divergence $\varDelta f$ as functions of correlation lag $\alpha$ and correlation strength $K$. The numerical simulation was done with the time difference $dt=0.01$ and the number of time steps $N_{step}=40000$. We chose the time interval from $t=360$ to $t'=400$ when calculating $I_{ij}(t,t')$. 
Each phase of the neurons is defined by $\theta_{j}=\arg(v_{j}+iw_{j})$.
From these measures, we can systematically determine the phase diagram even in a small-world graph that is  a non-Euclid complex system.
From these results, we found that synchronized region with $S_{sync}=0$ and $\varDelta f=0$ widely spreads at around $\alpha=0$ and $(\pm)\pi$.
Two main islands of chimera states with $0<S_{sync}/\log n\ll1$ and $\varDelta f >0$ are revealed in the synchronized sea at around $\alpha=\pm\pi/2$, which are shown as regions surrounded by white curves in Fig. 2(a).
On the one hand, the frequency is divergent at the middle in the left island at around $\alpha=-\pi/2$. On the other hand, the frequency almost stays $\varDelta f\varDelta t \lesssim 1$ in the right island chain.
The difference between the left and right chimera regions are further discussed later with reference to Fig. 3.

\begin{figure*}[hbt]
	\begin{center}
		\includegraphics[width=\linewidth]{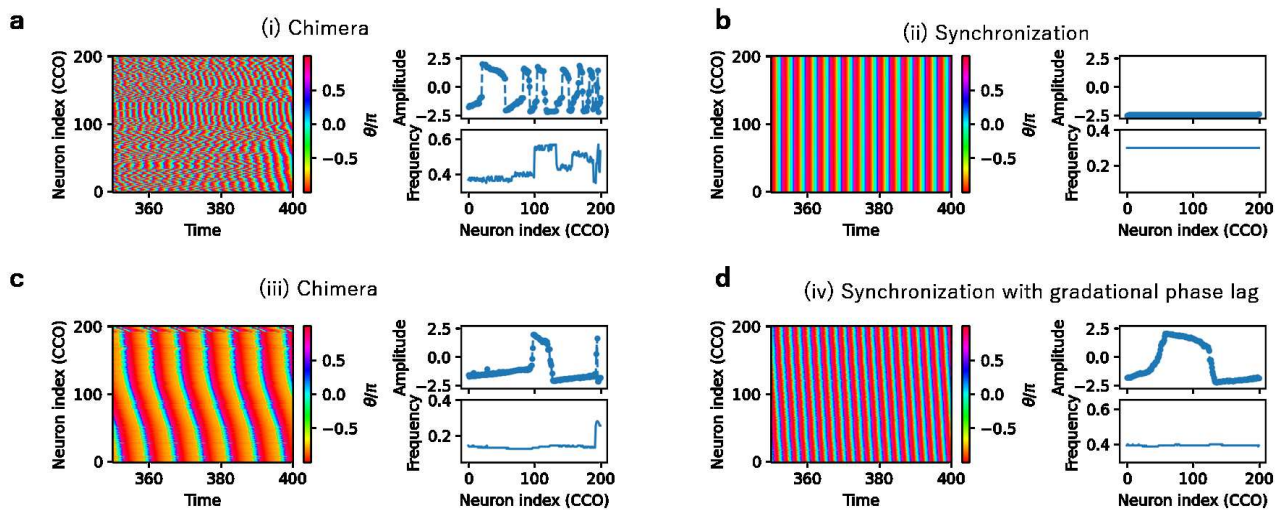}
	\end{center}
	\caption{Daynamics, snapshot, and mean frequency of reordered oscillators for several choices of the correlation lags depicted by stars on the synchronization phase diagram in Fig. 2. The correlation lags were set to $\alpha=-1.4, 0, 1.8 $ and $3$ for (a), (b), (c), and (d), respectively, with $K=8$. In each parameter setting, the left panel shows the phase dynamics of coupled FitzHugh-Nagumo neurons reordered by their synchronization groups calculated for an interval from $t=360$ to $t'=400$ and their phases at $t=400$. The right top panel is the snapshot of $v$-amplitudes at $t=400$. The right bottom panel is the mean spiking frequencies of reordered neurons during the same interval.}
	\label{fig3}
\end{figure*}

Figure \ref{fig2}(c) shows a total order parameter defined as $r=|\sum_j\exp(i\theta_j)|/N$, which characterizes the complete synchronization.
We can confirm that the synchronization sea at around $\alpha=0$ corresponds to the conventional synchronous phase. By comparing $r$ with our proposed measures, we also found another type of synchronized phase at around $\alpha=\pm \pi$, where the total order parameter $r$ does not develop.
As discussed below, this phase is a synchronized state with a gradational phase lag.
In the chimera regions, $r$ is small as expected.

Figure \ref{fig2}(d) shows temporal average of $g_0(t)$, where $g_0(t)=\int_0^\delta g(|D_{ij}|,t)d|D_{ij}|$, $D_{ij}=\exp(i\theta_i)-\exp(i\theta_j)$, and $g(|D_{ij}|,t)$ is a density function of $|D_{ij}|$ with a threshold for spacial coherence $\delta=0.01D_{max}$ \cite{kemeth_classification_2016}. This $g_0(t)$ corresponds to the relative size of the coherent component, and thus $g_0(t)=1$ means the complete synchronization. We found that this quantity successfully captures the synchronization phase, while, unfortunately, it cannot distinguish the other phases. Almost all regions except for the synchronized region at around $\alpha\sim0$ seem to be chimera states. The specific synchronized state with a gradational phase lag is overlooked by $g_0(t)$ as well as by $r$.

Figure \ref{fig3}(a)--(d) show the dynamics of a phase $\theta=\arctan(w/v)$ from time $t=350$ to $t=400$, amplitudes of $v$-components at $t=400$, and mean spiking frequencies $\omega$ of each neuron for different values of alpha with $K=8$, where these parameter points are shown as (i)--(iv) in Fig. \ref{fig2}(a). Note that the neuron indices are reordered firstly by the number of clique cluster to which they belonged and secondly by the values of the phase $\theta$ at $t=400$, which is referred to here as clique cluster ordering (CCO).
The ordering enables us to find a clear difference in dynamics among revealed phases.
First, as shown in Fig. \ref{fig3}(b), the synchronized region shows constant spiking frequency and amplitudes. 
Second, Fig. \ref{fig3}(d) shows  a constant spiking frequency and a gradual change in the amplitudes. This is the gradational phase lag synchronization state. Note that, the total order parameter cannot develop because of the gradational phase lag.
Furthermore, due to this phase lag, a measure $g_0$ index misleads us so that we mistake the gradational synchronized state for chimera states.
Third, in the chimera states shown in Fig. \ref{fig3}(a) and (c), some parts show that amplitudes of $v$ and spiking frequencies largely fluctuate, corresponding to the desynchronized parts. The other parts show almost constant spiking frequencies with amplitudes like a gradational synchronized state. These results indicate the coexistence of desynchronized and synchronized parts. The feature of the desynchronized parts is consistent with the behavior of the FHN model on the ring lattice structure \cite{omelchenko_ringFHN_2013,omelchenko_ringFHN_2015}. A phase gradation in the synchronized parts is a characteristic of this small-world network. Note that these chimera states have different characteristics: a dominant synchronous region in one (Fig. \ref{fig3}(c)), and  multiple synchronous regions in the other (Fig. \ref{fig3}(a)), which cause the aforementioned difference in $\varDelta f\varDelta t$.
In this way, the present measures allow us to visualize the synchronous phase from the viewpoint of discrete frequency synchrony.

\begin{figure*}[tb]
\begin{center}
\includegraphics[width=\linewidth]{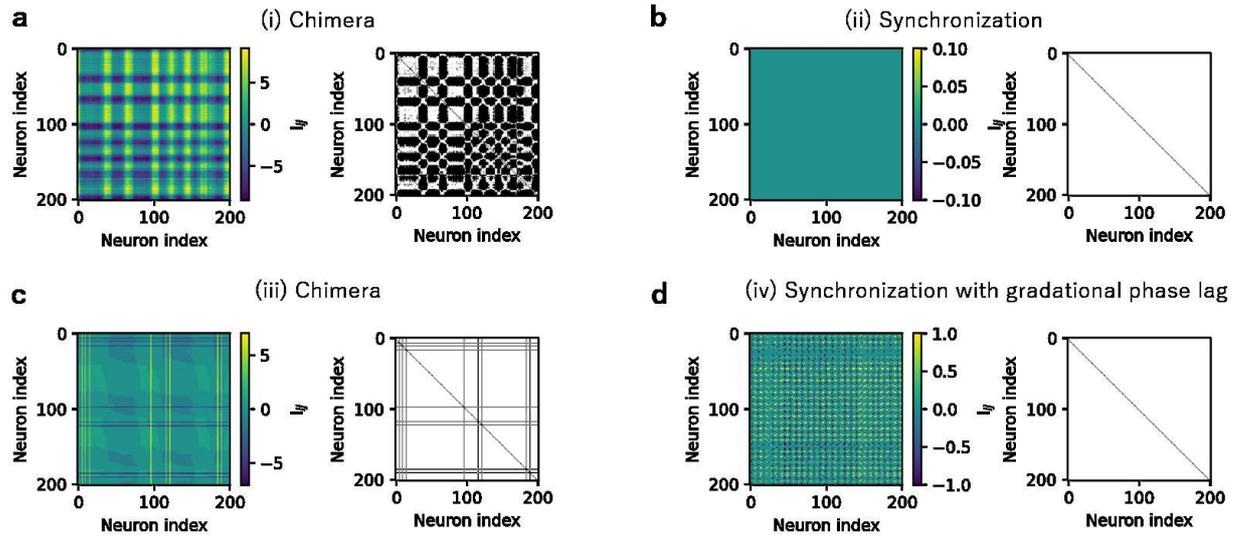}
\end{center}
\caption{False-color maps of pseudo-vorticity matrix and adjacent matrix of synchronization graph for coupled FitzHugh-Nagumo neuronal dynamics on a small-world network. We chose the correlation lags $\alpha=-1.4, 0, 1.8$, and $3$ for (a), (b), (c), and (d), respectively, with $K=8$. In each parameter setting, the left and right panels show the pseudo-vorticity matrix and adjacent matrix, respectively.}
\label{fig4}
\end{figure*}

We next show in Fig. \ref{fig4}(a)--(d) pseudo-vorticity matrix $I_{ij}$ (left) and the adjacency matrix $A_{ij}$ (right panel) of synchronization graph for four types of regions. We found that these four regions have distinct $I_{ij}$ and $A_{ij}$. 
Figure \ref{fig4}(b) and (d) show that two types of synchronized regions have $|I_{ij}| \le 1$ and complete graph of $A_{ij}$.
Note that the gradational phase lag synchronization has $|I_{ij}| = 1$, while complete in-phase synchronization only has $|I_{ij}| =0$.
For the two types of the chimera regions in Figs. \ref{fig4}(a) and (c), the magnitudes of $I_{ij}$ are large, and $A_{ij}$ are sparse and rather dense, respectively. This is because the chimeras in Fig. \ref{fig4}(a) and (c) are close to the desynchronized  and synchronized phases, respectively, which are consistent with the behavior in Figs. \ref{fig3}(a) and (c).
These characteristics of $I_{ij}$ provide a way to establish other good quantities in addition to $S_{sync}$ to distinguish the different phases.

\begin{figure}[tb]
\begin{center}
\includegraphics[width=8.8cm]{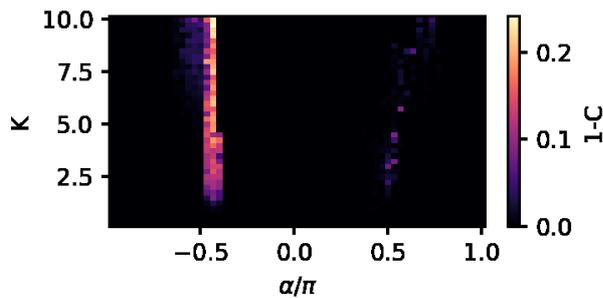}
\end{center}
\caption{False-color map of clustering coefficient $C$ for coupled FitzHugh-Nagumo neurons on a small-world network.}
\label{fig5}
\end{figure}
Figure~\ref{fig5} shows the clustering coefficient $C$ of the synchronization graph, and $1-C$ has very similar structure to the synchronization entropy. This is not so surprising because both scalers quantify the cliquishness of the graph from somewhat different points of view. The clustering coefficient may be useful for screening partial synchronizations in a large oscillatory system because it merely requires $O(n^3)$ calculation, which is much smaller than the cost to exactly solve the NP problem to obtain $S_{sync}$. We need further confirmation of the similarity between $C$ and $S_{sync}$ in different systems. We note that the temporal change in the graph structure includes the information on the synchronization dynamics. A comparison between the synchronization graph and the original network graph would also be interesting, which will be discussed elsewhere.

\section{Conclusions}
We have introduced a graphical description to classify complex synchronization dynamics in an oscillatory system based on a discrete analogue of frequency synchrony. To make the partial synchronization visible in a computer-friendly way, we have proposed an index matrix with integer elements from a topological viewpoint. Each element of the matrix represents the pseudo-vorticity generated by the phase difference accumulated by the time-evolution of a pair of oscillators in the system. From the matrix, we derive a synchronization graph that represents the pairwise synchronization in a graphical way. The synchronized clusters of oscillators are clarified by solving a clique cover problem of the graph. From the pseudo-vorticity matrix and the synchronization graph, we can also obtain the synchronization entropy and the frequency divergence of the system, which are scalar measures for partial synchronization. We apply our method to coupled neuronal oscillators on a small-world network based on the FitzHugh-Nagumo model. From the analysis, we confirmed that our proposed measures work well for classification of the phase dynamics with a phase diagram, where we found complete synchronization, gradational synchronization, and chimera states. 

In particular, the gradational synchronization and the related chimera states, which are hidden in the case of the original ordering of the oscillator index, are revealed by the reordering of the index based on the solution of the clique problem of the synchronization graph. 
The gradational synchronization missed in previous studies is one of the typical states, witch dominates a large portion of the parameter space. As well as revealing the frequency synchronized state with spacial-incoherence, we find the chimera state where synchronized and desynchronized states coexist accompanying the gradational phase lags.

We expect that our method will also work well to classify and quantify the partial synchronization of general discrete oscillatory systems appearing not only in numerical simulations but also in actual experiments. One interesting possibility is its application to electroencephalography (EEG) brain waves. An increase in partial synchrony on EEG signals for Parkinson's disease patients was found by the analysis based on pairwise order parameters~\cite{asher_connectivity_2021}, which quantify in-phase synchrony. With the proposed method, we can give a visualization of the synchronized clusters, and may find further partial synchrony with phase lags missed in the previous research. An another application can be found in functional analysis of neuromorphic devices. Adaptive and temporal changes in partial synchronization can be utilized for solving combinatorial optimization problems in photonic spiking neurons~\cite{inagaki_collective_2021}. Our method may contribute to revealing the mechanism in a visual and quantitative way.

\section*{Acknowledgments}
We thank Takahiro Inagaki for valuable discussion and Hiroyuki Tamura for his support during this research.

\appendix
\section{Generalization of synchronization entropy}
While we use Boltzmann-Gibbs-Shannon (BGS) entropy in Eq.~\eqref{eq:sync_entropy}, we have other choices for quantifying the uniformness of the size distribution in a partition of graph by cliques. In fact, we can obtain many entropy-like measures by using Schur-concave functions. A Schur-concave function gives a monotone of majorization, that is a binary relation of two stochastic states which determines whether one state can be mapped from the other state by a doubly stochastic linear map or not. BGS entropy is one of such Schur-concave functions. Here, we introduce another Schur-concave function as
\begin{align}
	S_{max}:=1-\max(\bm{p})=1-\frac{s_1}{n}.
\end{align}
$S_{max}$ is an intuitive measure of partial synchronization because $1-S_{max}$ is the largest cluster size, which contributes to a decrease in computing cost to solve the iterative clique problem if we don't need to know all the synchronized clusters of oscillators. It has the maximum $S_{max}=(n-1)/n$ for desynchronization. In synchronized cases, $S_{max}$ has the minimum $S_{max}=0$, whose value is the same as the synchronization entropy. 
We confirmed that $S_{max}$ show a qualitatively similar phase diagram as that of synchronization entropy $S_{sync}$ shown in Fig.~\ref{fig2} (data not shown).

\bibliography{manuscript}

\end{document}